\documentclass[epj,final]{svjour}
\usepackage{latexsym}
\usepackage{graphics}
\usepackage{array}
\usepackage{tabularx}
\usepackage{graphicx}
\usepackage{dcolumn}
\usepackage{bm}
\usepackage{amsmath}
\usepackage{multirow}

\newcolumntype{.}{D{.}{.}{-1}}
\newcolumntype{x}[2]{D{.}{.}{#1}p{#2}}

\begin{document}
\title{Neutron rich Carbon and Oxygen isotopes with an odd number of neutrons}
\subtitle{}
\author{B.~Laurent\inst{1} \and N.~Vinh Mau\inst{1,2}
}                     
%
%
\institute{CEA, DAM, DIF F-91297 Arpajon Cedex, France \and Institut de Physique Nucl\'{e}aire, IN2P3-CNRS, Universit\'{e} Paris-Sud, F-91406 Orsay Cedex, France }
\date{Received: date / Revised version: date}
%
\abstract{
We describe odd isotopes as formed of a core plus one neutron. We have
calculated the modification of single neutron energies brought by couplings of
the neutron with collective $2^+$ phonons in the cores. The results reproduce
very well the inversion of $2s$ and $1d_{5/2}$ shells in carbon isotopes up to
$^{19}$C while in oxygen isotopes the correction is also large but do not show
any inversion in agreement with experiments. The calculated energies are close
to the experimental ones in both series of isotopes except in $^{21}$C for the
$2s$ state for which our coupling is too weak.
} 
\maketitle
\section{Introduction}\label{sec:intro}
In the last decade the properties of carbon and oxygen isotopes have been
extensively studied experimentally and theoretically. They show very interesting
new behaviour when one compares with heavier nuclei. In particular carbon
isotopes offer a very interesting situation. Indeed  in lighter carbons,
including $^{12}$C, one notices an inversion of the $2s$ and $1d_{5/2}$ shells,
inversion which disappears when  N, the number of neutrons, is equal to 12 or
14. At contrary, in oxygen isotopes the normal sequence of $1d_{5/2}$ and $2s$
shells with the $1d_{5/2}$  lower than the $2s$ one is observed but neutron
energies for states closed to the Fermi surface are far from energies given by a
Saxon Woods potential or an Hartree Fock potential~\cite{peru12}. Therefore we are interested to see  what is the origin of
this inversion in carbon nuclei and how the situation is different in oxygen
nuclei. 

These nuclei have been the object of calculations mostly using shell models but
we present a different description based on the assumption of a two-body system
formed of a neutron added to  an inert core for which the two-body correlations
are introduced via a coupling between the neutron and a phonon of the core. The
corrections to the one-body neutron-core potential~\cite{vin95} due to this type
of coupling induces a modification of the single neutron energies which can be
easily calculated from the known properties of the phonons of the core.
In section~\ref{sec:model} we present our model and detail the calculation of
corrections due to two-body correlations introduced via neutron-phonon couplings. Then in section~\ref{sec:results} we present and discuss the
corresponding results obtained for carbon and oxygen isotopes. At last
section~\ref{sec:conclusion} is devoted to our conclusion.
      
\section{Description of isotopes with an odd number of neutrons and contribution
of Neutron-phonon couplings to the single neutron energies}\label{sec:model}
Assuming a neutron plus an inert core, the neutron states will be calculated
first using a Saxon-Woods one-body potential for the neutron-core interaction
written as~\cite{bohr}:

\begin{eqnarray}\label{eq:Vr}
V(r)=-V_0\left(f(r)-0.44 r_0^2({\bf l.s}) \frac{1}{r} \frac{df(r)}{dr}\right)
\end{eqnarray}
where
\begin{eqnarray}
f(r) & = & \left[1+exp\left(\frac{r-R_0}{a}\right)\right]^{-1}
\mbox{and}\label{eq:fr}\\
V_0  & = & \left(50.5-32.5~\frac{N-Z}{A}\right)~MeV\label{eq:V0}
\end{eqnarray}
with a parameter $a$ of 0.75~fm, larger than usual, to take account of the
diffuse surface of such light nuclei. The other parameters are $R_0=r_{0}
A^{1/3}$ with $r_0$=1.27~fm where $A$ is the nucleon number in the core.
These parameters were fitted in order to reproduce at best the neutron shells in
nuclei with $A>40$ where particle-phonon couplings are expected to have little
effect on single particle energies, therefore it can be considered as a good
phenomenological representation of the first order one-body potential.

To this first order potential we add the contribution due to two-body
correlations taken into account via a coupling between the neutron and a phonon
of the core. The corresponding modification of the potential is given in
references~\cite{vin95,vin69,pach02}. To calculate the corresponding modification
to the single neutron energies we follow reference~\cite{vin95}. This correction,
$\Delta\epsilon_n$ for the neutron in state $n$ is given by:  
 
\begin{eqnarray}
\Delta \epsilon_n=\Sigma_{N,\lambda} F_{N,\lambda}\left|\int \phi_n^*(r)
V_{0N}(r) \phi_\lambda (r) dr\right|^2\label{eq:De}
\end{eqnarray}
where:

- N and $\lambda$ mean respectively the phonons which are included in the
calculation and the state of the intermediate neutron. 

- $\phi_n$({\bf{r}}) and $\phi_\lambda$({\bf{r}}) are respectively the wave
functions calculated in the potential of Eq.~\ref{eq:Vr} for the neutron in
states $n$ and $\lambda$ with respective energies $\epsilon_n$ and
$\epsilon_\lambda$.

- V$_{0N}$ is the amplitude for exciting the phonon N of angular momentum L. As
in reference~\cite{vin95} we make a phenomenological approach and write V$_{0N}$ as:
\begin{eqnarray}
V_{0N}({\bf{r}})=\frac{1}{(2L+1)^{1/2}} \beta_L R_0 \frac{dU(r)}{dr}
Y_L^{M*}(\omega_r)\label{eq:V0N}
\end{eqnarray}
where U(r) is the central part of our Saxon Woods potential of
Eqs.~(\ref{eq:Vr}-\ref{eq:V0}) and $\beta_L$ is related to the B(EL) by:
\begin{eqnarray}
B(EL)=\left[\frac{3Z}{4\pi}R_c^2 \beta_L\right]^2~e^2 fm^4\label{eq:BEL}
\end{eqnarray}

- F$_{N,\lambda}$ is a coefficient given by:
\begin{eqnarray}
F_{N,\lambda}=(\frac{1-n_\lambda}{E_n-\epsilon_\lambda
-\omega_N})+(\frac{n_\lambda}{E_n-\epsilon_\lambda+\omega_N})\label{eq:FNL}
\end{eqnarray}
where $n_\lambda$ is the occupation number of the state $\lambda$ and $E_n$ is
the energy of the studied neutron after correction, which is then given by:  
\begin{eqnarray}
E_n=\epsilon_n+\Delta\epsilon _n\label{eq:En}
\end{eqnarray}

These equations have been solved by iteration to get the new energy E$_n$.

\section{Results for carbon and oxygen isotopes}\label{sec:results}
\subsection{Carbon isotopes}
All the results are given in table~\ref{tab:C} for the states closed to the
Fermi surface. We first concentrate on the energies of the $2s$ and $1d_{5/2}$
shells which present an unusual behaviour. We first calculate their energies in
our Saxon Woods potential of Eq.~\ref{eq:Vr}. We see that up to $^{20}$C the
$2s$ state is lower than the $1d_{5/2}$, then the order of the two shells is
inverted and the usual situation is recovered for $^{20}$C. The same situation
was already observed for $^{11}$Be in~\cite{vin95} when a pure Saxon Woods potential
was used. This property is clearly related to the radius of the neutron-core
potential. Indeed for smaller radii the centrifugal barrier is more efficient
and pushes up the d-states compared to the $2s$ one. However the calculated
energies are quite far from the experimental
ones~\cite{lan85,gui90,fau96,ajz80,ajz91,baz98,bau98,rid98,mar96,aud03,mad01_1,mad01_2,sta04,boh07,ele05,nak99,ele09,yam03}
which are also given in the table~\ref{tab:C_E}. Therefore we expect two body
correlations to be important as it was the 
case for the inversion of $1p_{1/2}$ and $2s$ shells in $^{11}$Be, $^{10}$Li and
$^{13}$Be. 
Indeed the cores that we are considering have $2^+$ states with a low excitation
energy and very large $ B(E2)$ due to a large collectivity. Such collective
$2^+$ states were shown to be responsible for a large modification of neutron
energies in a core of $^{10}$Be and to give most of the contribution to the
corrective term~\cite{vin95,gor04}. We give in Table II the energies and B(E2) for
the $2^+$ excited states that we have included in our calculation~\cite{thi04},
assuming that it gives most of the contribution. The results for cores of
$^{12}$C, $^{14}$C, $^{16}$C and $^{20}$C are given in the table~\ref{tab:C}
where we see that now the inversion of the two shells is amplified and
individual neutron energies for $2s$ and $1d_{5/2}$ shells are close to the
experimental ones. For a $^{18}$C-core we did not calculate the corrective term.
Indeed from the calculation in a SW potential and as given by experimental
energies the two shells are very closed, so that $^{18}$C is very likely a
mixture of $(2s)^2$ and 
$(1d_{5/2})^2$ configurations for the two last neutrons. It is indeed verified
when it is described as a core of $^{16}$C  plus two neutrons~\cite{bla}.
Therefore in our model it may not be taken as a core.

\begin{table}
  \caption{\label{tab:C} Results for carbon isotopes and comparison with
experimental values 
when they are known.}
   \centering
  \begin{tabular}{>{\raggedright}p{0.3cm}>{\raggedleft}p{0.5cm}||.|.|.|.|.}
         &                     & \multicolumn{1}{c}{\mbox{$2s_{1/2}$}} &
\multicolumn{1}{c|}{\mbox{$1d_{5/2}$}} & \multicolumn{1}{c|}{\mbox{$1d_{3/2}$}}
& \multicolumn{1}{c|}{\mbox{$2d_{5/2}$}} & \multicolumn{1}{c}{\mbox{$2d_{3/2}$}}
\\ \hline\hline
\multirow{4}{*}{$^{12}$C} & $\epsilon_n$      & -1.02      & -0.76      & 1.80     & 1.94       & 3.83\\
         & $\Delta \epsilon_n$ & -0.93        & -0.38      & -0.04      & -0.08    & -0.67 \\
         & E$_n$               & -1.95        & -1.14      & 1.76       & 1.86     & 3.16 \\
         & E$_{exp}$           & -1.86        & -1.12      &            & 1.74     & 3.3 \\ \hline
\multirow{4}{*}{$^{14}$C} &$\epsilon_n$       & -0.93      & -0.73      & 1.76     & 1.93       & 3.6 \\
         & $\Delta \epsilon_n$ & -0.21        & -0.06      & -0.01      & -0.01    & -0.16\\
         & E$_n$               & -1.13        & -0.79      & 1.75       & 1.92     & 3.44   \\
         & E$_{exp}$           & -1.22        & -0.48      &            &          & 3.47 \\ \hline
\multirow{4}{*}{$^{16}$C} &$\epsilon_n$       & -0.90      & -0.78      & 1.72     & 1.92       & 3.38 \\
         & $\Delta \epsilon_n$ & -0.37        &  0.        & -0.03      & 0.03     & -0.11 \\
         & E$_n$               & -1.27        & -0.78      & 1.69       & 1.95     & 3.27  \\
         & E$_{exp}$           & -0.73        & -0.48      &            &          & 3.2 \\ \hline
\multirow{4}{*}{$^{20}$C} &$\epsilon_n$       & -0.93      & -1.02      & 1.63     & 1.89       & 2.91 \\
         & $\Delta \epsilon_n$ & 0.37         & -0.18      & 0.05       & 0.01     & -0.02 \\
         & E$_n$               & -0.56        & -1.20      & 1.68       & 1.90     & 2.89  \\
         & E$_{exp}$           & 0.5          &            &            &          & 3. \\ 
  \end{tabular}
\end{table}

\begin{table}
  \caption{\label{tab:C_E} Energies and $B(E2)$ of the $2^+$ states used in our
calculation for the carbon cores}
   \centering
  \begin{tabular*}{0.45\textwidth}{@{\extracolsep{\fill}}l||.|.|.|.}
                    & \multicolumn{1}{c|}{\mbox{$^{12}$C}} &
\multicolumn{1}{c|}{\mbox{$^{14}$C}} & \multicolumn{1}{c|}{\mbox{$^{16}$C}} &
\multicolumn{1}{c}{\mbox{$^{20}$C}} \\ \hline\hline
E2 (MeV)            & 4.43   & 7.01    & 1.76   & 1.62 \\ \hline
B(E2) (e$^2$fm$^2$) & 39.7     & 18.7      & 13.      & 18.4 \\
  \end{tabular*}
\end{table}

The same calculation has been done for $1d_{3/2}$, $2d_{5/2}$ and $2d_{3/2}$
shells and results are given in table~\ref{tab:C}. 
For higher neutron states the corrections due to the couplings are negligible as
expected. Our $2d_{5/2}$ reproduces very well the second $5/2^+$ excited state
in $^{13}$C~\cite{aud03} while there is no experimental
$3/2^+$ state corresponding to our $1d_{3/2}$ state. This state is very close
to the second $5/2^+$ and perhaps not easily detected experimentally. However our $2d_{3/2}$ is
in great agreement with the known high $3/2^+$ state in all our systems
even though the corrective term is very small.

For a $^{20}$C core we recover the usual order with the $1d_{5/2}$ lower than
the $2s$, therefore $^{20}$C may be assumed to have a closed $1d_{5/2}$ shell.
However for the $2s$ state the neutron-phonon coupling is not strong enough to
make the state unbound as it should. In this case we may expect a further
contribution to the potential of a term corresponding to the excitation of
collective pairing vibrations in the $N-2$ and $N+2$ systems~\cite{vin69}, therefore in
$^{18}$C and $^{22}$C. Indeed in particular in $^{18}$C it has been shown that
the wave functions are mixture of several configurations for the two last
neutrons~\cite{bla}. This type of contribution was found long ago to have a
noticeable contribution in $^{40}$Ca~\cite{bou74}.

As a summary we can say that, even though we have restricted the neutron-phonon
coupling to the low energy $2^+$ phonon, we have been able to reproduce the
inversion of $2s$-$1d_{5/2}$ shells in the lightest carbon isotopes (up to
$^{20}$C) and could get reasonable agreement with experimental neutron energies
for all calculated states.

\subsection{Oxygen isotopes}
We have performed the same calculations for oxygen isotopes assumed to be
described as an inert core plus one neutron. We have considered cores of
$^{14}$O, $^{16}$O, $^{20}$O and $^{25}$O.$^{18}$O may not be assumed as a core.
Indeed because of the proximity of the $2s$ and $1d_{5/2}$ shells, the wave
function is a mixture of two configurations with the two last neutrons in one of
these two shells. As for carbons we have calculated the contribution to the
neutron energies of neutron-phonon couplings due to the low energy $2^+$ excited
state of the core. The energies and B(E2) are taken from measurements~\cite{hof08,ram01}
and given in table~\ref{tab:O_E}. 
The results of our calculation are reported in table~\ref{tab:O} for $2s$,
$1d_{5/2}$ and $1d_{3/2}$ neutron states together 
with the experimental energies~\cite{hof08,hof09,bel01,jur07,cor04}.

\begin{table}
  \caption{\label{tab:O}Results for oxygen isotopes and comparison with
experimental values when they are 
known~\cite{hof08,hof09,ram09}.}
   \centering
  \begin{tabular*}{0.45\textwidth}{@{\extracolsep{\fill}}lr||.|.|.}
         &                     & \multicolumn{1}{c|}{\mbox{$2s_{1/2}$}} &
\multicolumn{1}{c|}{\mbox{$1d_{5/2}$}} & \multicolumn{1}{c}{\mbox{$1d_{3/2}$}}
\\ \hline\hline
\multirow{4}{*}{$^{14}$O} &$\epsilon_n$       & -3.5     & -4.92    & 1.53
    \\
         & $\Delta \epsilon_n$ & -0.89      & -0.79      & -0.27      \\
         & E$_n$               & -3.90        & -5.02      & 0.99       \\
         & E$_{exp}$           & -4.50        & -6.4       & 1.54       \\
\hline
\multirow{4}{*}{$^{16}$O} &$\epsilon_n$       & -3.2       & -4.52      & 1.37  
    \\
         & $\Delta \epsilon_n$ & -0.61        & -0.54      & -0.5       \\
         & E$_n$               & -3.81        & -4.05      & 0.87       \\
         & E$_{exp}$           & -3.3         & -4.14      & 0.94       \\
\hline
\multirow{4}{*}{$^{22}$O} &$\epsilon_n$       & -2.85      & -4.1       & 0.65  
    \\
         & $\Delta \epsilon_n$ & 0.12         & 0.15       & 0.26       \\
         & E$_n$               & -2.73        & -3.95      & 0.91       \\
         & E$_{exp}$           & -2.74        & -3.8       &            \\
\hline 
\multirow{4}{*}{$^{24}$O} &$\epsilon_n$       & -2.82      & -4.08      & 0.38  
    \\
         & $\Delta \epsilon_n$ &              &            &            \\
         & E$_n$               &              &            &            \\
         & E$_{exp}$           & -4.09        & -5.05      & 0.77       \\ 
  \end{tabular*}
\end{table}

\begin{table}
  \caption{\label{tab:O_E} Energies and $B(E2)$ of the $2^+$ states in the
oxygen cores used in our calculation.}
   \centering
  \begin{tabular*}{0.45\textwidth}{@{\extracolsep{\fill}}l||.|.|.}
                    & \multicolumn{1}{c|}{\mbox{$^{14}$O}} &
\multicolumn{1}{c|}{\mbox{$^{16}$O}} & \multicolumn{1}{c}{\mbox{$^{22}$O}}\\
\hline\hline
E2 (MeV)            & 6.59   & 6.92   & 3.19 \\ \hline
B(E2) (e$^2$fm$^2$) & 43.         & 40.6     & 21.   \\
  \end{tabular*}
\end{table}

We see first that we have no inversion of the $2s$ and $1d_{5/2}$ shells for any
of the isotopes in agreement with experiments. On the other side the neutron-phonon couplings are quite important and put the calculated energies close to
the experimental ones. As in carbon isotopes, for higher states and already for
the $2d_{5/2}$ and $2d_{3/2}$ states, these corrections are negligible
($<$0.1~MeV). For $^{25}$O we do not know any $2^+$ state then could not
calculate the coupling term.  
However we note that to obtain the experimental energy, in particular for the
$2s$ state, we need a large corrective term what suggests 
that there is a low $2^+$ state in $^{24}$O with a large B(E2). This is plausible since in $^{24}$O the $1d_{5/2}$ and $2s$ shells are closed to each
other and a $2^+$ state can be constructed by coupling two neutrons of the
$1d_{5/2}$ shell to $J^{\pi}=2^+$. The excitation energy of this state should be
then close to the low $2^+$ state of $^{18}$O known at 1.98~MeV.

\section{Conclusions}\label{sec:conclusion}
To conclude we were able to understand and reproduce the properties of single
neutron states in carbon and oxygen isotopes with an odd number 
of neutrons. We have shown the importance of two body correlations which are
treated in our work as neutron-phonon couplings reduced to 
coupling with the lowest $2^+$ state in the core. In particular we could explain
that for carbon isotopes the $2s$ and $1d_{5/2}$ shells are 
inverted up to $^{20}$C as shown experimentally. For oxygen isotopes no such
inversion appears in our calculations as required by experiments. For both
series of nuclei we get neutron energies closed to experimental values except
for $^{21}$C where the coupling of the neutron with the $2^+$ phonon of $^{20}$C
is not strong enough to get the measured unbound energy of the $2s$ state even
though it reduces the discrepancy between calculation and measurement. 

Apart the corrections on single neutron energies, particle-phonon couplings
induce a modification of the wave functions which are now a mixture 
of one single state with configurations where one neutron is coupled to
phonons~\cite{vin95}. One, sometimes, makes a parallel between this mixture of
complex configurations and a deformation of the corresponding state as it is
explicitly used in sulfur studies~\cite{che12}. This equivalence was already clear
in the case of $^{11}$Be where the inversion of $1/2^+$ and $1/2^-$ was shown to
be due to two-body correlations in a calculation similar to the
present one~\cite{vin95} or to a deformation of the one-body average potential~\cite{nun96_1,nun96_2} due
to the presence of the low $2^+$ state in $^{10}$Be which was considered as a
vibrational state in the first case and a rotational state in the second one.

\section*{Acknowledgement}
We would like to thank J. P. Ebran, G.
Blanchon, S. Peru-Desenfants and N. Pillet for their interest in our work and for many valuable discussions about
results.

\bibliographystyle{/cea/home/pn/laurentb/biblio/unsrtlinuxmodif}
\bibliography{bib_epja}

\end{document}